\documentclass[12pt]{article}

\pdfoutput=1
\usepackage{jheppub}
 \usepackage{etoolbox}
    \makeatletter
    \patchcmd{\maketitle}{\@fpheader}{}{}{}
    \makeatother
\usepackage{url}
\usepackage{color}
\usepackage{fancyhdr}
\usepackage{amsthm}
\usepackage{float}
\usepackage{subfigure}
\usepackage{enumerate}

\newcommand{\be}{\begin{equation}}
\newcommand{\ee}{\end{equation}}
\newcommand{\bal}{\begin{aligned}}
\newcommand{\eal}{\end{aligned}}

\def\tr{\mathop {{\rm \, Tr}} \nolimits}         
\def\Tr{\tr}

 at10pt

\title{Zero Modes and Entanglement Entropy}

\author[a,b]{Yasaman K. Yazdi}

\affiliation[a]{Perimeter Institute for Theoretical Physics, 31 Caroline St. N., Waterloo ON, N2L 2Y5, Canada}
\affiliation[b]{Department of Physics and Astronomy, University of Waterloo, Waterloo ON, N2L 3G1, Canada}

\emailAdd{yyazdi@perimeterinstitute.ca}

\begin{document}

\abstract{

Ultraviolet divergences are widely discussed in studies of entanglement entropy. Also present, but much less understood, are infrared divergences due to zero modes in the field theory. In this note, we discuss the importance of carefully handling zero modes in entanglement entropy. We give an explicit example for a chain of harmonic oscillators  in 1D, where a mass regulator is necessary to avoid an infrared divergence due to a zero mode. We also comment on a surprising contribution of the zero mode to the UV-scaling of the entanglement entropy.}

\pagestyle{plain}
\thispagestyle{empty}

\maketitle

\flushbottom

\section{Introduction}
Entanglement entropy has been an important topic in various fields of theoretical physics for some time, and interest continues to grow in this deep and useful concept.  Especially in the quantum gravity community, many believe that the key insight that will help us connect quantum mechanics and general relativity, will come from entanglement entropy. Some works in this direction include \cite{TJ} and \cite{VanRam}. Related to these motivations is the question of black hole entropy and whether all of it or most of it is entanglement entropy. The earliest works conjecturing that this might be the case are by R. Sorkin \cite{raf, RDS2}. Many others have thought about this as well (e.g. \cite{srednicki}), but we do not have a final answer to this question yet.

In quantum field theory and AdS/CFT many important theorems involving entanglement entropy have been proved \cite{Cardy, Cardy2, Ryu} such as strong subadditivity \cite{Ryu2}. Also in condensed matter theory, important applications of entanglement entropy include investigating topological order \cite{Kitaev} as well as properties of Fermi surfaces \cite{Swingle}, where quantum phase transitions are characterized by the entanglement entropy of the system.

While the ultraviolet divergences of entanglement entropy are widely discussed, infrared divergences are much less studied. A simple example of an infrared divergence in $1+1$D occurs in a massless theory of a chain of harmonic oscillators or scalar field on an interval with periodic boundary conditions \cite{RDSlaw, Casini2, Guifre, Unruh, SV}. Interesting new work is also being done where IR divergences arise, such as in entanglement entropy of excited states in conformal perturbation theory  \cite{Antony}, entanglement in bandlimited quantum field theory \cite{Achim}, and others \cite{Casini, Casini1, Pad}.  It is therefore worth understanding more rigorously in simple systems.

In this work we study in detail the divergence of the entanglement entropy in a simple theory as a result of a zero mode. Zero modes, analogous to free particles, do not have normalizable ground states. Theories that possess  zero modes,  such as the massless scalar field on a circle (spacetime cylinder), therefore do not have well-defined ground states \cite{Abd}. One may still define a ground state for such a theory by either ignoring the zero mode solution \emph{ad hoc}, or else by somehow regulating it. If it is included in the theory and not regulated, it can lead to infrared divergences, for example in the entanglement entropy. We study an example of this below.

Casini and Huerta \cite{Casini, Casini1} have found infrared divergences in the massless limit of a scalar field theory on a finite interval within an infinite line. In this work, the system we study is an interval of a chain of harmonic oscillators within a \emph{finite} chain with periodic boundary conditions (i.e., a circle).  It is interesting that the infrared divergence found by \cite{Casini, Casini1} on the infinite line is a double logarithm of the mass, while we find a single logarithm divergence on the finite circle. Additionally, in the present paper we pinpoint the source of the infrared divergence as a giant eigenvalue of an operator derived from the Lagrangian. This giant eigenvalue additionally has the interesting property that it also contributes to the scaling of the entanglement entropy with the UV cutoff in our finite system. Thus this eigenvalue cannot simply be discarded, as it contributes to the UV scaling of the entanglement entropy, and must therefore be regulated appropriately.

\section{Entropy of Oscillators}
We consider a chain of harmonic oscillators, with nearest-neighbour couplings. To find the entanglement entropy associated to a subchain (a connected subset of the full chain) of this system, we follow the procedure laid out in \cite{RDS2}, which we now review.

The Lagrangian for our chain of oscillators is

\be
\mathcal{L}=\frac{1}{2}\left(\sum_{N=1}^{N_{max}} \hat{\dot{q}}_N^2-\sum_{N, M=1}^{N_{max}} V_{MN}\,\hat{q}_N \,\hat{q}_M\right)=\frac{1}{2}\sum_{N=1}^{N_{max}} [\hat{\dot{q}}_N^2-m^2 \hat{q}_N^2-k(\hat{q}_{N+1}-\hat{q}_N)^2],
\label{Lpho}
\ee
where k is the coupling strength between the oscillators, and in terms of the spatial UV cutoff $a$, $k=1/a^2$ \cite{gold}. We define the positive, symmetric matrix $W$ using the potential of this Lagrangian:

\be
W_{MA}{W^A}_N=V_{MN},
\label{sqrt}
\ee
where the symmetric, positive definite metric $G_{MN}$ and its inverse $G^{MN}$ given by $G^{MP}G_{PN}={\delta^M}_N$, is used to raise the index in \eqref{sqrt}. $G_{MN}$ is a metric on the configuration space of the coupled harmonic oscillators. 
Now we consider the division of our chain of harmonic oscillators into a subchain whose oscillators will be labelled with Greek indices, and the remainder of the chain whose oscillators will be labelled with Latin indices. It is convenient to rewrite $W$ in terms of blocks referring to these two divisions:

\[W_{AB}= \left( \begin{array}{cc}
W_{ab} & W_{a\beta}  \\
W_{\alpha b} & W_{\alpha\beta} \end{array} \right).\] 

Following the convention of \cite{RDS2}, the inverse of $W_{AB}$ will be expressed as 
\[W^{AB}= \left( \begin{array}{cc}
W^{ab} & W^{a\beta}  \\
W^{\alpha b} & W^{\alpha\beta} \end{array} \right),\] 
 and the inverse of each block will be expressed with tildes (for example $\widetilde{W}^{ab}$ is the inverse of $W_{ab}$).
It was shown in \cite{RDS2} that when $\rho$ is the density matrix for the vacuum state, the reduced density matrix $\rho_{red}$ associated to a subchain of oscillators (say, the Latin-indexed ones) can be expressed in terms of these blocks we have just defined, as

\be
\rho_{red}(q^a, q'^b)=\sqrt{det(\widetilde{W}_{ab})} \exp[-\frac{1}{2}W_{ab}(q^a q^b+q'^a q'^b)]\exp[\frac{1}{4}\widetilde{W}^{\alpha\beta}W_{\alpha a}W_{\beta b}(q+q')^a(q+q')^b].
\ee

 Furthermore, it was shown that the entropy $S=-\Tr\rho_{red}\ln\rho_{red}$ can be expressed in terms of the eigenvalues $\lambda_n$ of the operator $\Lambda^a_b\equiv W^{ac} W_{c\alpha}\widetilde{W}^{\alpha\beta}W_{\beta b}$, as

\be
S=\sum_n\{\ln(\frac{1}{2}\sqrt{\lambda_n})+\sqrt{1+\lambda_n}\ln(\sqrt{1+1/\lambda_n}+1/\sqrt{\lambda_n})\}.
\label{lam}
\ee

An alternative method to compute the entanglement entropy involves a matrix $C = \sqrt{XP}$, where $X_{ij}$ are the field correlators at sites $i$ and $j$, and $P_{ij}$ are the conjugate momentum correlators at sites i and j in the region corresponding to the reduced density matrix.  The entanglement entropies calculated using that method should match those found in this paper\footnote{See e.g. Section 2.2.1 of \cite{Casini} for further details.}. Below we compute the entanglement entropy associated to a shorter subchain within a longer chain of oscillators with periodic boundary conditions, using the formula \eqref{lam}. We will study the zero mode in this model and show that it leads to an infrared divergence. We single out the source of the divergence as a giant eigenvalue in the spectrum of $\Lambda$. We also consider the entanglement entropy associated to a subchain within a longer chain of oscillators with one fixed boundary. We show that no infrared divergences arise when setting the mass to zero in this case, since there are no zero modes.

\subsection{Periodic Boundary Conditions}
\label{uvgiantsec}
We consider the Lagrangian \eqref{Lpho}, with periodic boundary conditions $\hat{q}_{N+1}=\hat{q}_1$. If we set the mass to zero, the entropy diverges logarithmically, as we will show. If we set $m$ to a small but finite number, the entropy is finite and obeys the expected asymptotic form for $a\rightarrow 0$ of logarithmic scaling with the UV cutoff \cite{Cardy, Cardy2, Ryu}
\be
S\sim\frac{1}{3}\ln[{L \sin({\pi\ell/L})/\pi a}]+c_1,
\label{spmz}
\ee
 where $\ell$ and $L$ are the physical lengths (number of oscillators times the spacing $a$) of the subchain and total chain respectively, and $c_1$ is a non-universal constant whose exact form is known for a few systems (eg. \cite{BQ1, BQ2}). In the limit that the length of the smaller subchain is much shorter than the length of the full chain ($\frac{\ell}{L}\rightarrow 0$), the entropy simplifies to

\be
S\sim\frac{1}{3}\ln(\ell/a)+c_1.
\label{spmzs}
\ee

 The entropy in \eqref{spmz} or \eqref{spmzs} is only well-defined with respect to an overall vacuum state. In referring to this result for the massless theory on a circle, one has to be careful to address the fact that this theory does not have a well-defined ground state due to the zero mode. Before moving on to a discussion of this zero mode, we first verify that by regulating the zero mode with a small mass, the result of \eqref{spmz} can indeed be obtained.

 \begin{figure}[H]
\begin{center}
\includegraphics[width=0.8\textwidth]{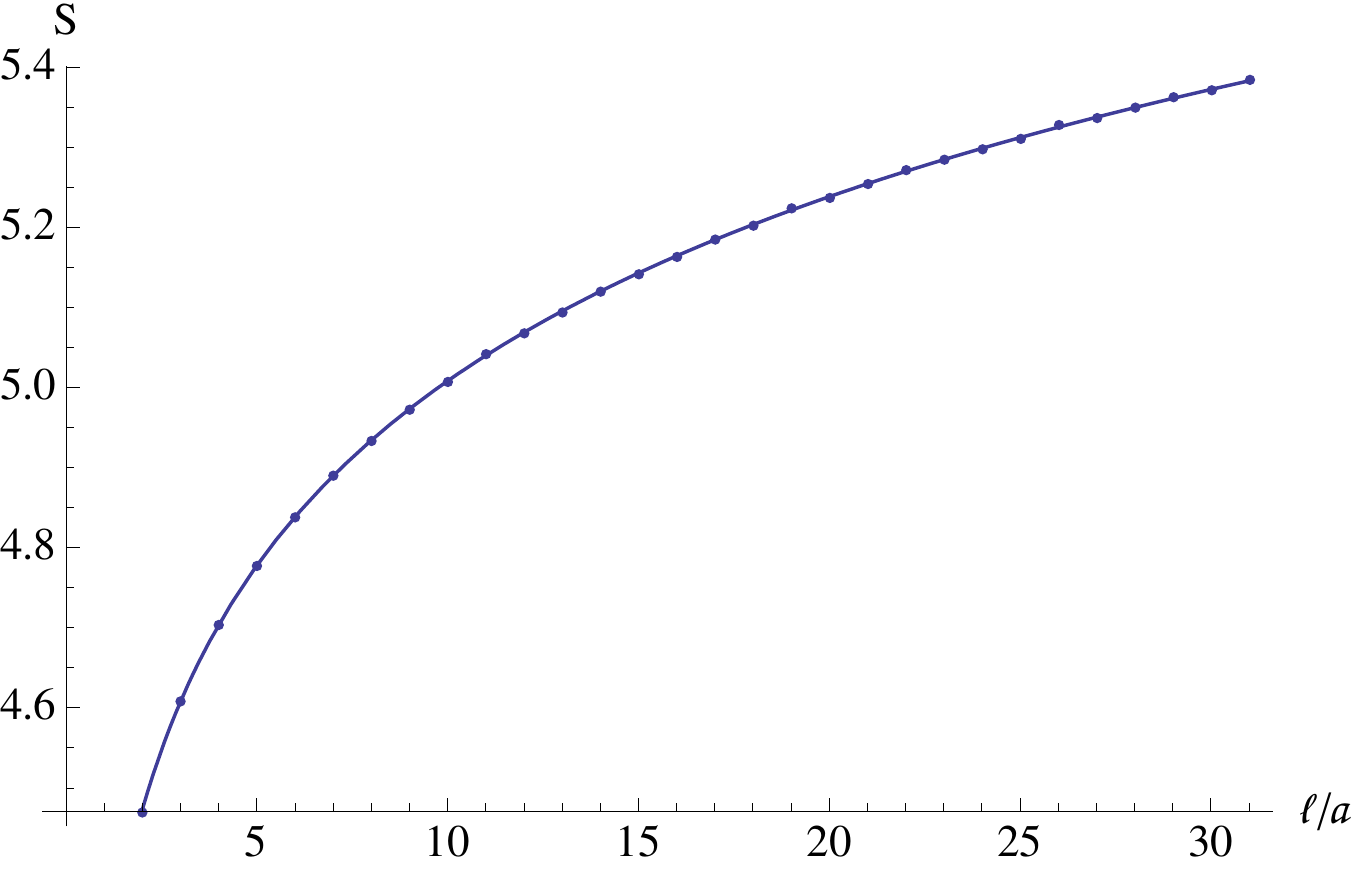}
\caption{S vs. $\mathcal{\ell}/a$ for a chain of harmonic oscillators with periodic boundary conditions.}
\label{spho}
\end{center}
\end{figure}

We regulate the zero mode with a small mass $m^2=10^{-6}$ and set $k= 10^6$. Our chain of oscillators contains 500 oscillators. We hold $k$ fixed and vary $\ell$. The result is shown in Figure \ref{spho}, where the solid curve is the function $S=b_1\ln({\sin({\pi\ell/L})})+c_1$ being fit to the data. The best fit parameters are $b_1=0.3337$ and $c_1=5.9316$, in agreement with \eqref{spmz}.

\begin{figure}[H]
\begin{center}
\includegraphics[width=0.8\textwidth]{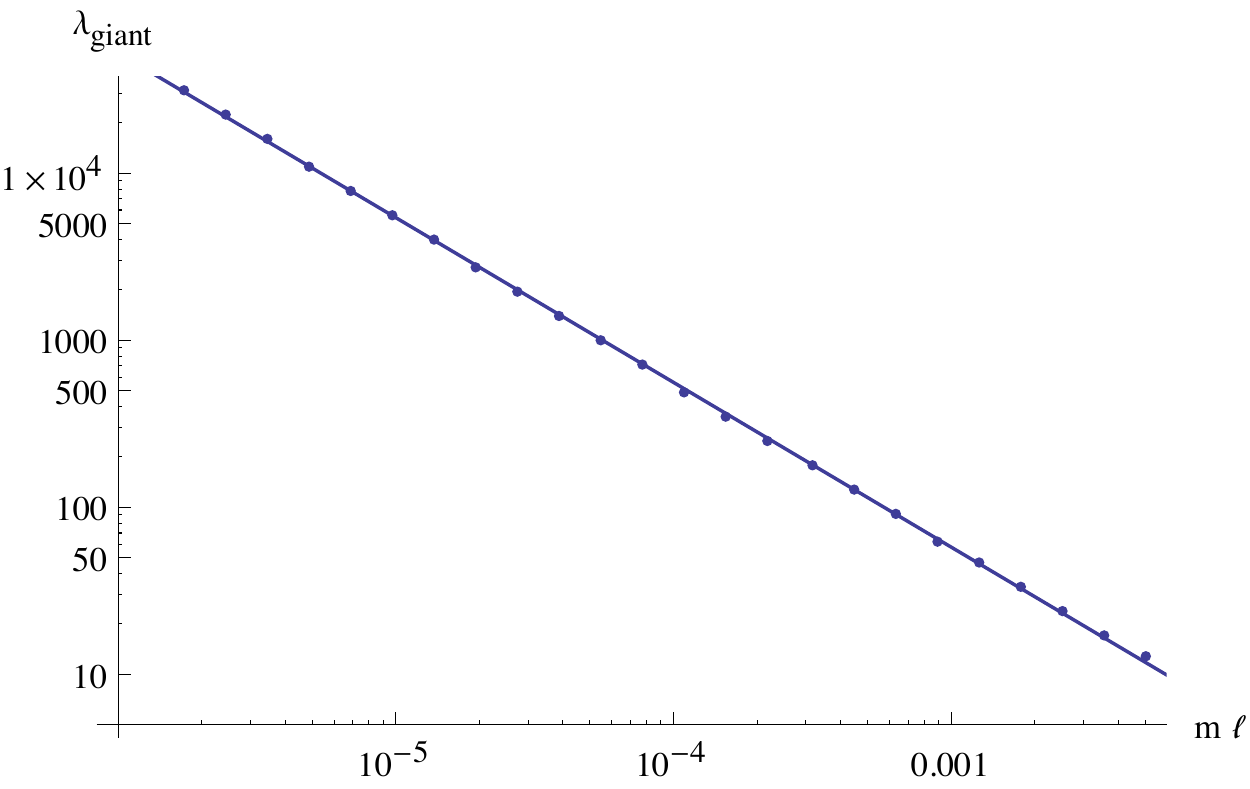}
\caption{$\lambda_{giant}$ vs. $m \mathcal{\ell}$ on a log-log scale for a subchain of 10 harmonic oscillators within a longer chain of 500 oscillators with periodic boundary conditions.}
\label{lg}
\end{center}
\end{figure}

\begin{figure}[H]
\begin{center}
\includegraphics[width=.8\textwidth]{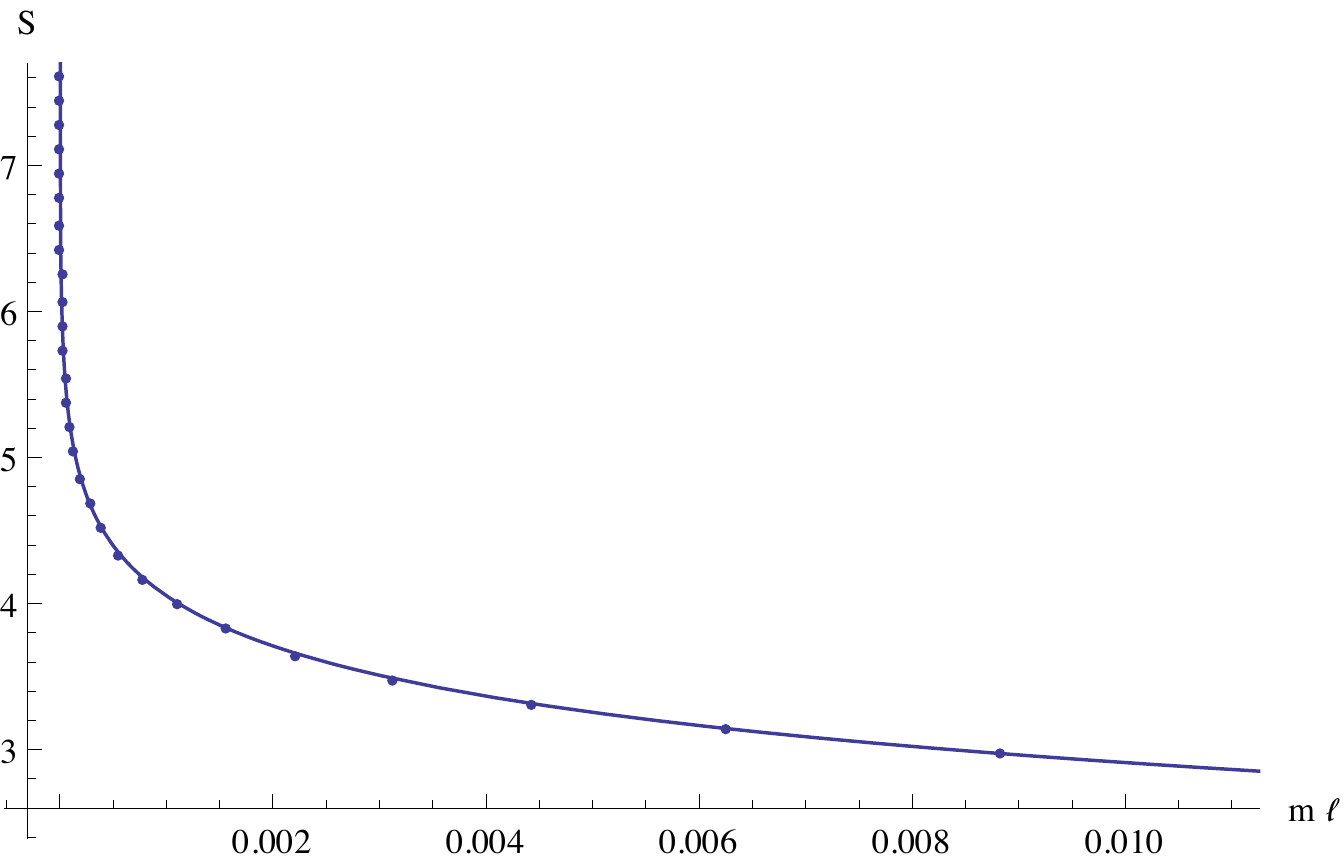}
\caption{S vs. m$\mathcal{\ell}$ for a subchain of 50 harmonic oscillators within a longer chain of 500 oscillators with periodic boundary conditions.}
\label{zm50}\
\includegraphics[width=.8\textwidth]{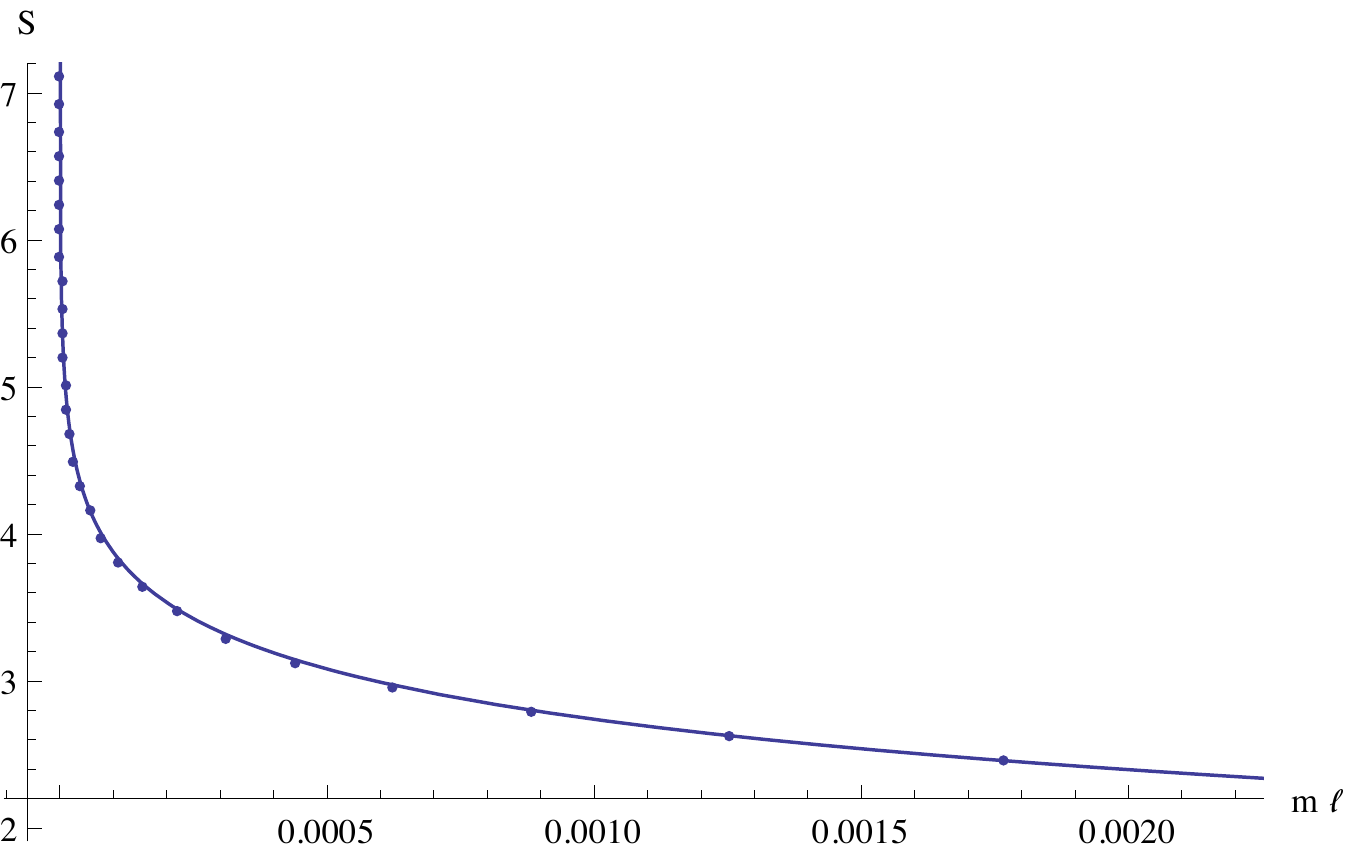}
\caption{S vs. m$\mathcal{\ell}$ for a subchain of 10 harmonic oscillators within a longer chain of 500 oscillators with periodic boundary conditions.}
\label{zm10}
\end{center}
\end{figure}

The contribution of the regulated zero mode to the entanglement entropy can be seen in the spectrum of $\Lambda$. It contributes a giant eigenvalue (relative to the other eigenvalues). A sample spectrum for a subchain of 10 oscillators within a chain of 500 oscillators, with $k=10^6$ and $m^2=3\times10^{-8}$ is $\{31400, 0.470, 0.0321, 2.03\times10^{-3}, 9.82\times10^{-5}, 
 3.49\times10^{-6}, 8.88\times10^{-8}, 1.53\times10^{-9}, 1.66\times10^{-11}, 
 2.91\times10^{-13}\}$. The size of the giant eigenvalue is inversely proportional to the size of the mass regulator ($\lambda_{giant}\propto m^{-1}$). This leads to a power law divergence in the giant eigenvalue as $m \rightarrow 0$, as illustrated in Figure \ref{lg} for the 10-subchain. The form of \eqref{lam} then suggests that we should expect a logarithmic divergence of the entropy with the size of this mass regulator. This is indeed what we find. Figures \ref{zm50} and \ref{zm10} show the results for varying $m$ in the limit $m\rightarrow0$, while holding all other variables and parameters fixed. We considered subchains of $50$ and $10$ oscillators, within a chain of $500$ oscillators. The 50-subchain result fits $S=b_2 \ln(m\mathcal{\ell})+c_2$ with $b_2=-0.496$ and $c_2=0.624$, while the 10-subchain result fits the same function with $b_2=-0.494$ and $c_2=-0.674$. The infrared divergence is therefore logarithmic, with a universal coefficient of $\sim -\frac{1}{2}$. 
 
\begin{figure}[H]
\begin{center}
\includegraphics[width=.8\textwidth]{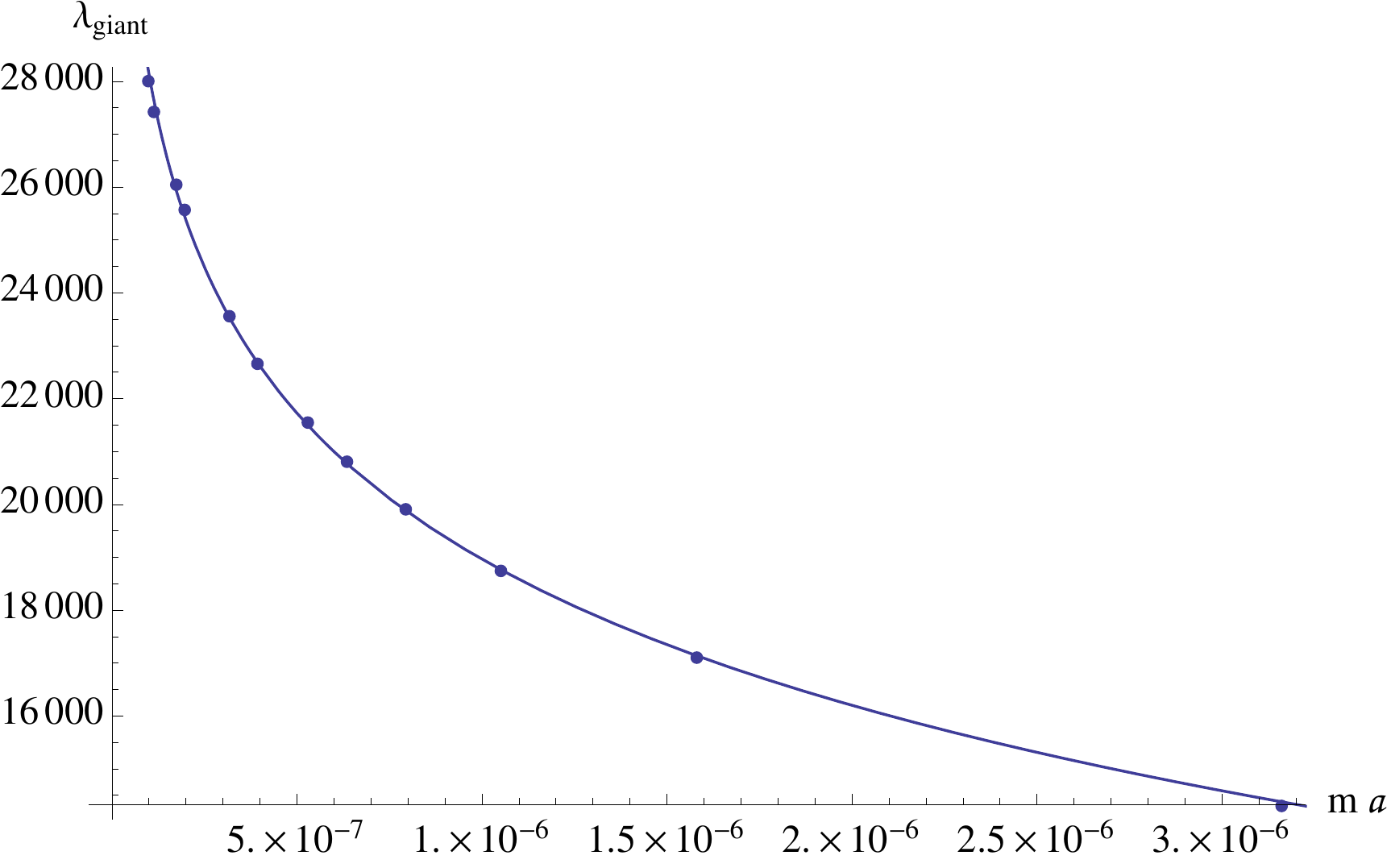}
\caption{$S$ vs. $m a$ for a subchain of length $\ell=\frac{5}{100\sqrt{10}}$ within a longer chain of length $L=10 \ell$, with fixed mass $m^2=10^{-6}$ and periodic boundary conditions.}
\label{uvgiant}\
\end{center}
\end{figure}

Surprisingly, the size of the giant eigenvalue also has a dependence on the UV cutoff. This dependence is logarithmic: $\lambda_{giant}=b_3 \ln(m a )+c_3$ (keeping $m$ fixed), with $b_3$ and $c_3$ being non-universal constants that are inversely proportional to the length of the entire chain $L$. Figure \ref{uvgiant}  shows the result for $\ell=\frac{5}{100\sqrt{10}}$ \footnote{In the numerical code for the computations of Figures 5 and 6, our initial value for the square of the oscillator spacing was $a^2=1/k=1/10^5$ or $a=1/(100\sqrt{10})$, and $\ell=n a$ (with $n=5$).  As stated after \eqref{Lpho}, $a$ is related to the constant $k$ by $k = 1/a^2$, which is why $a$ contains the $\sqrt{10}$ in its expression. For each subsequent data point we divide $a^2$ by $y^2$ and multiply $n$ by $y$, where $y$ is an integer, such that we keep $\ell$ fixed. }, $L=10 \ell$, and $m^2=10^{-6}$ and is fit by $\lambda_{giant}=b_3 \ln(m a )+c_3$ with $b_3=-3985$ and $c_3=-36093$.

\begin{figure}[h]
\begin{center}
\includegraphics[width=0.9\textwidth]{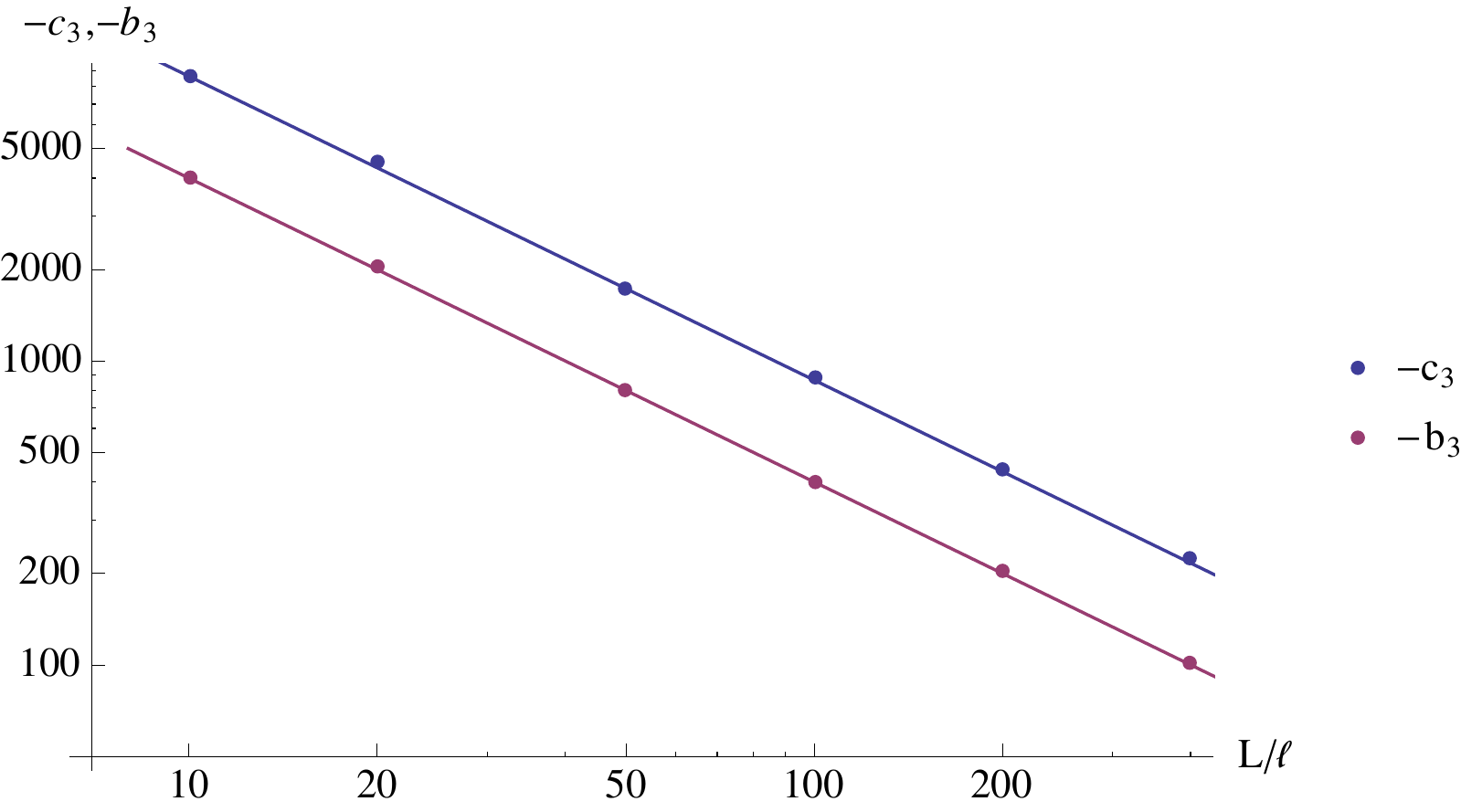}
\caption{$-b_3$ and $-c_3$ vs. $L/\ell$ on a log-log scale for a subchain of length $\ell=\frac{5}{100\sqrt{10}}$, $m^2=10^{-6}$, and periodic boundary conditions.}
\label{bc3}
\end{center}
\end{figure}

Figure \ref{bc3} shows the scaling of $b_3$ and $c_3$ with the length of the entire chain $L$. We set $\ell=\frac{5}{100\sqrt{10}}$, $m^2=10^{-6}$, and impose periodic boundary conditions on the chain of oscillators. We set $L/\ell$ to different values, and for each of these values find $b_3$ and $c_3$ from the UV scaling. The fits correspond to $-b_3=d_1\frac{\ell}{L}$ with $d_1\sim 4.0\times 10^4$, and $-c_3=d_2\frac{\ell}{L}$ with  $d_2\sim 8.6\times 10^4$. Therefore, we see that indeed $b_3 \propto L^{-1}$ and $c_3  \propto L^{-1} $, resulting in the UV dependence of the zero mode becoming sub-leading compared to the total contribution of all the other eigenvalues in the limit $L\rightarrow \infty$. This is consistent with other works, such as \cite{Casini, Casini1}, that do not see a significant contribution from the zero mode to the UV scaling in the limit $L\rightarrow\infty$.

\subsection{One Fixed Boundary}
The chain of harmonic oscillators with Dirichlet boundary conditions at one end no longer possesses a zero mode (translation symmetry is broken), and as a result of this we do not expect there to be any infrared divergences for the massless theory. The Lagrangian for the oscillators is again \eqref{Lpho}, but with boundary condition $\hat{q}_{N+1}=0$. Since a mass regulator is no longer needed, we set $m=0$, and $k=10^6$ as before.

The expected asymptotic form for the entanglement entropy, with $a\rightarrow 0$, is \cite{Cardy, Cardy2, Ryu}
\be
S\sim\frac{1}{6}\ln({L \sin({\pi\ell/L})/\pi a})+c_1.
\label{s1b}
\ee

\begin{figure}[H]
\begin{center}
\includegraphics[width=0.8\textwidth]{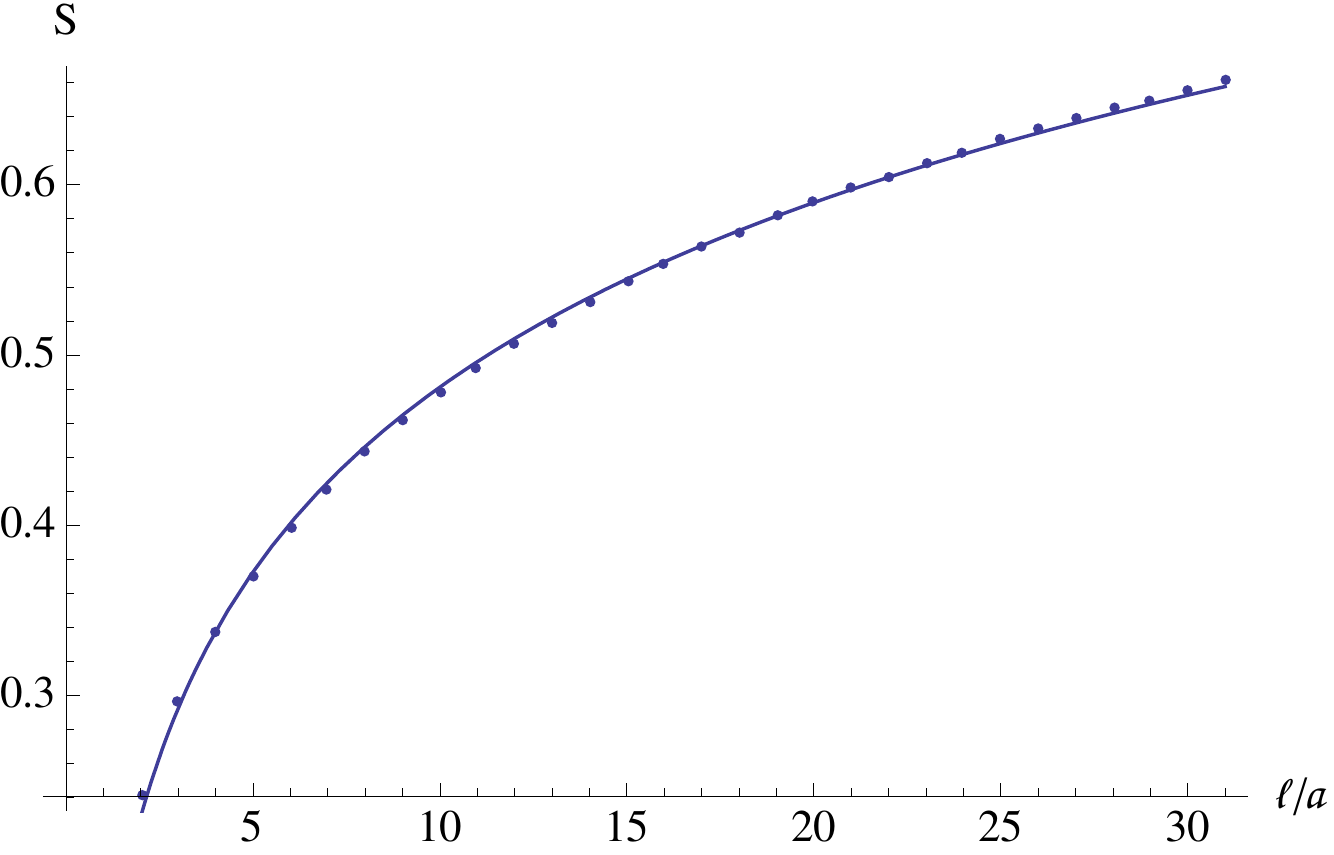}
\caption{S vs. $\mathcal{\ell}/a$ for a chain of harmonic oscillators with one fixed boundary.}
\label{scg}
\end{center}
\end{figure}

 Our result for the harmonic oscillators with one fixed boundary is shown in Figure \ref{scg}, and $S=b_1\ln({\sin({\pi\ell/L})})+c_1$ is fit to this data. The best fit parameters are $b_1=0.1567$ and $c_1=0.9150$, in agreement with \eqref{s1b}. Also, as expected, there is no longer a giant eigenvalue in the spectrum of $\Lambda$. A sample spectrum for a subchain of 10 oscillators within a chain of 500 oscillators, with $k=10^6$ and $m=0$ is $\{0.738, 7.72\times 10^{-3}, 5.98\times 10^{-5}, 2.56\times10^{-7}, 6.06\times10^{-10}, 
 6.98\times10^{-13}, 6.66\times10^{-14}, 
 2.44\times10^{-14}, -1.62\times10^{-15}, -1.98\times10^{-15}\}$.
 
\begin{figure}[H]
\begin{center}
\includegraphics[width=0.68\textwidth]{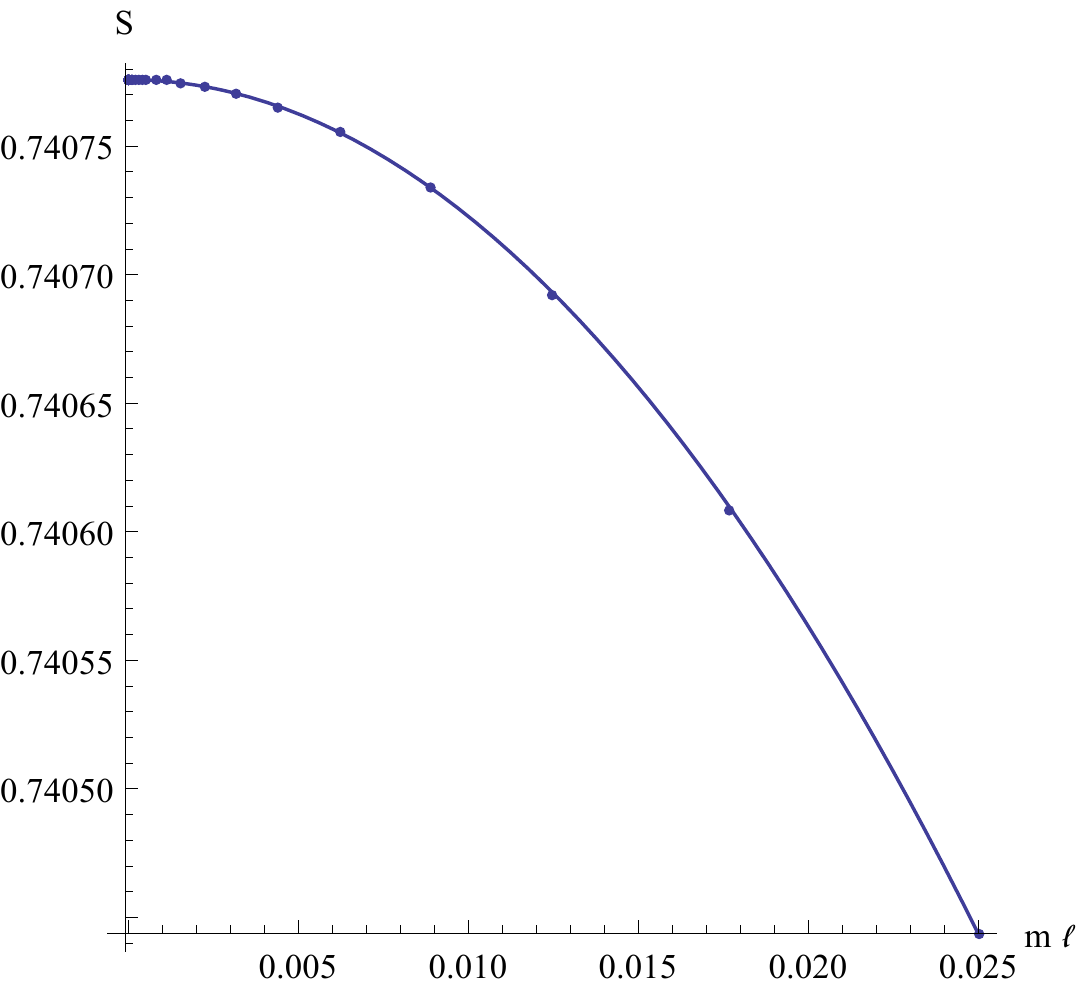}
\caption{S vs. $m \mathcal{\ell}$ for a subchain of 50 harmonic oscillators within a longer chain of 500 oscillators with one fixed boundary.}
\label{1bnzm}
\end{center}
\end{figure}

We have just established that there is no infrared divergence for the chain with one fixed boundary, but we will nevertheless examine more closely the small mass limit of this case. Figure \ref{1bnzm} shows the result for varying $m$ in its limit $m\rightarrow0$, while holding all other variables and parameters fixed. The subchain consists of $50$  oscillators, within a longer chain of $500$ oscillators. The entropy remains finite for all small $m$ including $m=0$. The data fits $S=b_4 ~ (m\mathcal{\ell})^2+c_4$ with best fit parameters $b_4=-0.532$ and $c_4=0.741$. 

\section{Conclusions}
We have demonstrated through our specific example, the importance of correctly regulating a theory containing zero modes in order to get physically meaningful scalings for entanglement entropy.  We saw that the unregulated theory has an IR divergence in the entanglement entropy, and thus obscures all interesting UV physics.  Finally, we demonstrated the resolution of this problem by introducing a mass regulator to the theory, revealing the interesting behaviour of the entropy. We also found that the zero mode has a dependence on the UV cutoff which makes a significant contribution to the entanglement entropy for finite $L$ and becomes sub-leading in the limit $L\rightarrow \infty$.

Another example of an IR regulation of the zero mode is described in \cite{yas}. There, the regulation is done by placing a massless scalar field in a causal diamond instead of infinite Minkowski space.

\section{Acknowledgements}
We thank Rafael Sorkin for suggesting this calculation, and for discussions. We also thank Niayesh Afshordi, Anton van Niekerk, and Horacio Casini for helpful discussions. This research was supported in part by Perimeter Institute for Theoretical Physics. Research at Perimeter Institute is supported by the Government of Canada through the Department of Innovation, Science and Economic Development Canada and by the Province of Ontario through the Ministry of Research, Innovation and Science.


\begin{thebibliography}{25}

\bibitem{TJ}
T. Jacobson, \emph{Entanglement Equilibrium and the Einstein Equation}, (2015) arXiv:1505.04753v3 [gr-qc].
\bibitem{VanRam}
N. Lashkari, M. B. McDermott, M. Van Raamsdonk, \emph{Gravitational Dynamics From Entanglement ``Thermodynamics"}, JHEP 1404 (2014) 195,  	arXiv:1308.3716 [hep-th].

 
\bibitem{raf}
  R. D. Sorkin, \emph{On the Entropy of the Vacuum Outside a Horizon}, in Tenth International Conference on General Relativity and Gravitation (held Padova, 4-9 July, 1983), Contributed Papers, vol. 2, pp. 734736.  (1983)  	arXiv:1402.3589 [gr-qc].
 
\bibitem{RDS2}Luca Bombelli, Rabinder K. Koul, Joohan Lee, and Rafael D. Sorkin, \emph{A Quantum Source of Entropy for Black Holes}, Phys. Rev. D34, 373-383 (1986).

 \bibitem{srednicki}
 M. Srednicki, \emph{Entropy and Area}, PRL 71 (1993) 666Ð669.
 
\bibitem{Cardy} P. Calabrese and J. Cardy, \emph{Entanglement Entropy and Conformal Field Theory}, J. Phys. A 42, 504005 (2009) arXiv:0905.4013 [cond-mat.stat-mech].
\bibitem{Cardy2}
P. Calabrese and J. Cardy, \emph{Entanglement Entropy and Quantum Field Theory}, J. Stat. Mech (2004), no. P06002.
\bibitem{Ryu}
S. Ryu and T. Takayanagi, \emph{Holographic Derivation of Entanglement Entropy from AdS/CFT}, Phys. Rev. Lett. (2006), no. 96.

\bibitem{Ryu2}
M. Headrick and T. Takayanagi, \emph{A Holographic Proof of the Strong Subadditivity of Entanglement Entropy}, Phys.Rev.D76:106013, (2007) arXiv:0704.3719 [hep-th].

\bibitem{Kitaev}
A. Kitaev, J. Preskill, \emph{Topological Entanglement Entropy},  Phys.Rev.Lett. 96 (2006) 110404, (2006) , arXiv:hep-th/0510092.

\bibitem{Swingle}
B. Swingle, \emph{Entanglement Entropy and the Fermi Surface}, (2010),  arXiv:0908.1724 [cond-mat.str-el].

\bibitem{RDSlaw}
A. Chandran, C. Laumann, and R. D. Sorkin, \emph{When is an Area Law not an Area Law?}, (2015), arXiv:1511.02996 [hep-th].

\bibitem{Casini2}
H. Casini and M. Huerta, \emph{A Finite Entanglement Entropy and the c-theorem}, Phys.Lett. B600 (2004) 142-150, arXiv:hep-th/0405111.

\bibitem{Guifre}
G. Evenbly and G. Vidal, \emph{Entanglement Renormalization in Free Bosonic Systems: Real-Space Versus Momentum-Space Renormalization Group Transforms}, New J. Phys. 12, 025007 (2010), arXiv:0801.2449 [quant-ph].

\bibitem{Unruh}
W. G. Unruh, \emph{Comment on ``Proof of the Quantum Bound on Specific Entropy for Free Fields"}, Phys. Rev. D 42, 3596, 1990.

\bibitem{SV}
T. He, J. M. Magan, and S. Vandoren, \emph{Entanglement Entropy of Periodic Sublattices}, Phys. Rev. B 95, 035130 (2017), arXiv:1607.07462.

\bibitem{Antony}
A. J. Speranza, \emph{Entanglement Entropy of Excited States in Conformal Perturbation Theory and the Einstein Equation},  (2016) arXiv:1602.01380 [hep-th].

\bibitem{Achim}
J. Pye, W. Donnelly, and A. Kempf, \emph{Locality and Entanglement in Bandlimited Quantum Field Theory},  	Phys. Rev. D 92, 105022, (2015)	arXiv:1508.05953 [quant-ph].

\bibitem{Casini}
  H. Casini and M. Huerta, \emph{Entanglement Entropy in Free Quantum Field Theory}, J.Phys. A42 (2009) 504007, arXiv:0905.2562 [hep-th].
  
\bibitem{Casini1}
  H. Casini and M. Huerta, \emph{Entanglement and Alpha Entropies for a Massive Scalar Field in Two Dimensions},  	J.Stat.Mech.0512:P12012, (2005), arXiv:cond-mat/0511014.

\bibitem{Pad}
K. Mallayya, R. Tibrewala, S. Shankaranarayanan, and T. Padmanabhan, \emph{Zero Modes and Divergence of Entanglement Entropy}, (2014) arXiv:1404.2079 [hep-th].

\bibitem{Abd}
  E. Abdalla, M. Abdalla, and K. Rothe,
\emph{Non-perturbative Methods in 2 Dimensional Quantum Field Theory}.  World Scientic Pub Co Inc, (1991).



\bibitem{gold}
H. Goldstein, \emph{Classical Mechanics}, Addison-Wesley, (1950).
\bibitem{BQ1}
B.-Q. Jin and V. E. Korepin, \emph{Quantum Spin Chain, Toeplitz Determinants and Fisher-Hartwig Conjecture}, J. Stat. Phys. 116, 79 (2004) [quant-ph/0304108].
\bibitem{BQ2}
A. R. Its, B.-Q. Jin, and V. E. Korepin, \emph{Entanglement in XY Spin Chain}, J. Phys. A 38, 2975 (2005) [quant-ph/0409027].
\bibitem{yas}
N. Afshordi, M. Buck, F. Dowker, D. Rideout, R. D. Sorkin, and Y. K. Yazdi, \emph{A Ground State for the Causal Diamond in 2 Dimensions}, JHEP 10, 088 (2012), no. 10, arXiv:1207.7101.

\end{thebibliography}
\end{document}